\documentclass[11pt,a4paper]{article}

\usepackage{cite}
\usepackage{amsmath, amsthm, amssymb,slashed}
\usepackage{ifpdf}
\ifpdf
  \usepackage[pdftex]{graphicx}
  \usepackage{epstopdf}
\else
  \usepackage[dvips]{graphicx}
\fi
\usepackage{fancyhdr}
\usepackage{mathbbol}
\usepackage{url}
\usepackage{color}
\usepackage{bbm}
\usepackage{dsfont}
\usepackage{bm}
\usepackage{amsfonts}

\def\be{\begin{equation}}
\def\ee{\end{equation}}
\def\bea{\begin{eqnarray}}
\def\eea{\end{eqnarray}}

\parskip=\baselineskip

\def\1{{\bf 1}}
\def\2{{\bf 2}}
\def\3{{\bf 3}}
\def\4{{\bf 4}}

\font\teneurm=eurm10 \font\seveneurm=eurm7 \font\fiveeurm=eurm5
\newfam\eurmfam
\textfont\eurmfam=\teneurm \scriptfont\eurmfam=\seveneurm
\scriptscriptfont\eurmfam=\fiveeurm

\font\teneusm=eusm10 \font\seveneusm=eusm7 \font\fiveeusm=eusm5
\newfam\eusmfam
\textfont\eusmfam=\teneusm \scriptfont\eusmfam=\seveneusm
\scriptscriptfont\eusmfam=\fiveeusm

\font\tencmmib=cmmib10 \skewchar\tencmmib='177
\font\sevencmmib=cmmib7 \skewchar\sevencmmib='177
\font\fivecmmib=cmmib5 \skewchar\fivecmmib='177
\newfam\cmmibfam
\textfont\cmmibfam=\tencmmib \scriptfont\cmmibfam=\sevencmmib
\scriptscriptfont\cmmibfam=\fivecmmib

\begin{document}
\begin{titlepage}
\begin{flushright}

\end{flushright}
\vskip 1.5in
\begin{center}
{\bf\Large{Constraints from precision measurements  on the hadron-molecule  \\\vspace{0.5truecm}   interpretation of $X,Y,Z$ resonances
}}
\vskip 0.9cm {AD Polosa$^{*,\P}$} \vskip 0.05in {\small{ \textit{$^*$Dipartimento di Fisica, Sapienza Universit\`a di Roma, Piazzale Aldo Moro 2, I-00185 Roma, Italy}\vskip -.4cm
{\textit{$^\P$INFN Sezione di Roma, Piazzale Aldo Moro 2, I-00185 Roma, Italy }}}
}
\end{center}
\begin{center}
\makeatletter
\tiny\@date
\makeatother
\end{center}
\begin{abstract}
The precise measurement of  binding energy, total width and $DD\pi$ branching fraction, constrains, in a model independent way, the loosely bound hadron molecule interpretation of  $X(3872)$. A full discernment is not yet possible on the basis of present data. Similar considerations might be extended to all those $Y,Z$ resonances susceptible of a molecular description.
\end{abstract}
%
\end{titlepage}

The literature on the loosely bound molecule interpretation of  $X(3872)$ and  several other $Y,Z$ states, is very large, see for example the references quoted in~\cite{rass}, and it influenced a rather  wide consensus, also among experimentalists, on its validity as the correct and  economical picture for the description of most of $X,Y,Z$ states --  even though it is the less appealing picture in terms of any new understanding in strong interactions dynamics. 

The well known fact that the formation of prompt  $X$s  from  $D\bar D^*$ pairs coalescing into  unstable bound states is  not a natural option in high energy proton-proton collisions~\cite{bigna}, even when final state multi-body interactions are taken into account~\cite{rescatt},  should be the problem to be solved in order to  explain production data, prior to any statement on loosely bound molecules phenomenology. However some authors believe  that the issues on high energy production will eventually be solved resorting to strong interaction rescattering mechanisms of the sort of those proposed in~\cite{artobrat}. 

On the experimental side, data on the production of (anti)deuteron at high transverse momenta --  $p_T\gtrsim 8$~GeV -- in LHC collisions, especially at ALICE~\cite{rass}, might have a strong discriminating meaning when compared to the high $p_T$ prompt production cross section of $X$ as measured by CMS, see {\it e.g.}~\cite{cms}.  Some simple extrapolations of available low-$p_T$ data on anti-deuteron production at ALICE seem to suggest that we cannot expect any significant production rate at high $p_T$, whereas the $X$ is  copiously observed at CMS with $p_T$ hadron cuts as large as $p_T\gtrsim 10$~GeV.  If confirmed in future analyses,  the lack of anti-deuterons at high $p_T$ cuts  would naturally  speak against any deuteron-like interpretation of $X$.

In this brief report I discuss, with a standard~\footnote{The quantum mechanics formulas presented in this note are standard. I review them concisely in the context of $X,Y,Z$ phenomenology. } and model-independent argument, how a precise experimental determination 
of the binding energy ${\mathcal E}$ (\mbox{${\mathcal E}=m_D+m_{D^*}-m_X\gtrsim0$}), total width $\Gamma_X$ and ${\cal B}(X\to DD\pi)$ branching fraction, would further constrain the loosely bound hadron molecule picture of  $X(3872)$. Similar considerations might equally be extended to $Y,Z$ molecular candidates. For the time being we only know that $\Gamma_X\lesssim 1.2$~MeV and ${\cal B}(X\to DD\pi)>32\%$ with $m_X=3871.69\pm0.17$~MeV -- precise determination of $D$ and $D^*$ masses enter as well in the determination of the binding energy ${\mathcal E}$. 

Consider the amplitude for the elastic process $\alpha\to \beta $. Using the $1/\sqrt{2E}$ wave function normalization for in and out particles and the appropriate $1/(2\pi)^3$ factors, one can write the scattering amplitude~\footnote{$d\sigma/d\Omega=|f|^2$} as~\cite{wein}
\be
f(\alpha\to \beta)=-\frac{1}{8\pi E}A_{\beta\alpha}
\ee
where $E$ is the energy in  the $\alpha$ state. 

Suppose now that the initial and final states are two spinless particles and  there is a pole dominance in the transition amplitude $A_{\beta\alpha}$,  due to some intermediate  state $c$. In the particular case of the (elastic) scattering of two hadrons $a$ and $b$ we have $\alpha=\beta=ab$ and
\be
f(ab\to c\to ab)=-\frac{1}{8\pi E}\,g^2\,\frac{1}{(p_a+p_b)^2-m_c^2}
\label{pole}
\ee
where $g$ (which has the dimensions of a mass) is proportional to the strength of the attractive force responsible for their interaction. Let $c$ be an (unstable) bound state of $a$ and $b$ with binding energy ${\mathcal E}\to 0^+$. We have
\be
m_c\simeq m_a+m_b-{\mathcal E}
\label{mc}
\ee
and, in the non-relativistic approximation of slowly recoiling $a,b$
\be
E\simeq m_a+m_b
\label{ene}
\ee
whereas we expand
\be
(p_a+p_b)^2\simeq (m_a+m_b+T)^2
\label{mom}
\ee
where $T=\bm p_a^2/2m_a+\bm p_b^2/2m_b=\bm p^2/2m$ is the, supposedly small, recoil kinetic energy in the center-of-mass of the $ab$ pair, where $\bm p_a=-\bm p_b$, and $m$ is the $ab$ reduced mass. 
Plugging (\ref{mc}), (\ref{ene}) and (\ref{mom}) in~(\ref{pole}) we have
\be
f(ab\to c\to ab)\simeq-\frac{1}{16\pi (m_a+m_b)^2}\,g^2\,\frac{1}{{\mathcal E}+T}
\label{pole2}
\ee
neglecting  small ${\mathcal E}^2$ and $T^2$ terms.

In the non-relativistic quantum mechanics description of resonant low energy scattering, see~\cite{land}, two slow particles $(kR\ll 1)$  interacting through an attractive potential $U$ (of range $R$), with a superficial discrete level at $-{\mathcal E}$ ($|{\mathcal E}|\ll |U|$), have a universal  elastic scattering amplitude
\be
f(ab\to ab)=-\frac{1}{\sqrt{2m }} \frac{\sqrt{{\mathcal E}}-i\sqrt{T}}{{\mathcal E}+T}
\label{nrqm}
\ee
which is independent of the details of the potential $U$, being affected only by the value of the resonant level at  $-{\mathcal E}$. A brief proof of this formula is provided below.

The kinetic energy $T$ in the latter formula, coincides with the total energy $T=E>0$ in  region $II$ ($r>R$) outside the potential range, where free motion takes place with wave function $\chi(r)_{II}\sim\sin(kr+\delta_0)$ -- given that we assume $kR\ll1$, $\chi(r)_{II}$ varies slowly as $r\to 0$.  
Because of the slow variation of $\chi_{II}$, the matching condition $(\chi^\prime/\chi)_{II}=(\chi^\prime/\chi)_{I}$, to be taken at some $r^*>0$ point, could formally be computed at $r^*=0$.  Therefore  we obtain $(\chi^\prime/\chi)_{II}=k\cot\delta_0$. 

Within region $I$ ($r<R$), the Schr\"odinger equation will not depend explicitly on energy, as $U\gg |{\mathcal E}|$, and the boundary condition will not depend on the total energy either. Given the independency on $E$ in region $I$, let us choose to be at the stationary state. In correspondence of  a generic boundary value $r^*$, we have that $\chi_I=A\sin k_I r^*\equiv Be^{-\kappa r^*}$ and $\kappa=\sqrt{2m{\mathcal E}}/\hslash$. We choose the latter form for the boundary condition so to be independent on $U$ and $r^*$: $(\chi^\prime/\chi)_{I}=-\kappa$. 

Since $k$ in region $II$ is $k=\sqrt{2mT}/\hslash$,  the boundary condition at $r^*$ is  $\cot\delta_0=-\sqrt{{\mathcal E}/T}$. The latter formula can be used in the S-wave scattering amplitude~\footnote{
\be
f(\alpha\to \beta)=\frac{1}{k(\cot\delta_0-i)}\\
\ee
}  leading eventually to Eq.~(\ref{nrqm}). 

From these considerations, the determination of the scattering length $a$ follows~\footnote{
At very low energies, $k\sim 0$, in region $II$ we actually have to solve $\chi^{\prime\prime}_{II}=0$, which has the solution $\chi_{II}\sim (r-a)$ (infinitely long-wave-limit of a $\sin$ function). Therefore it also holds that ${\rm lim}_{k\to 0} k\cot\delta_0={\rm lim}_{k\to 0} (\chi^\prime/\chi)_{II} =-1/a$, which defines the scattering length $a$. From the matching condition found above, $1/a=\kappa$ so that $a=\hslash/\sqrt{2m{\mathcal E}}$ -- as it is also found discussing the Low equation as described in~\cite{WeiMec} -- and $\sigma=4\pi a^2=2\pi\hslash^2/m{\mathcal E}$.}  and a comparison between Eq.~(\ref{nrqm}) and~(\ref{pole2}),  the latter encoding  the dependency on the interaction force in the  $g$ coupling, leads to the following  relation 
\be
{\mathcal E}\simeq\frac{g^4}{512\pi^2}\frac{m^5}{(m_am_b)^4}
\label{lform}
\ee
which, again, is independent on the details of $U$. 

In the case of  the $X(3872)$ loosely bound molecule, we should take $m_a=m_{D}$, $m_b=m_{D^*}$ and define the strong coupling $g$ through~\footnote{Since $X$ has positive charge conjugation, the final state is $|f\rangle=(|D^0\bar D^{0*}\rangle +|\bar D^0 D^{0*}\rangle)/\sqrt{2}$. When extracting $g$ defined in~(\ref{coupl})  from data a factor of $\sqrt{2}$ has to be included: $g\to \sqrt{2}g$.} 
\be
\langle D^0\bar{D}^{0*}(\epsilon,q)|X(\lambda,P)\rangle=g\,\lambda\cdot \epsilon^*
\label{coupl}
\ee

In principle $g^2$ is derived from the $\Gamma(X\to D\bar D^*)$ decay width, which is $\Gamma\sim g^2 \Phi$ if particles are considered to be spinless ($\Phi$ being the two-body decay phase space).
Taking into account the spin of  $D^*$ and $X$, one should rather substitute in~(\ref{lform})
\be
g^2\to g^2\frac{1}{3}\left(2 + \frac{(m_X^2+m_{D^*}^2-m_{D}^2)^2}{4m_X^2m_{D^*}^2}\right)
\label{corr}
\ee 
which, however, turns out to be numerically $\simeq g^2$. The actual value of $g$  is extracted from data on the branching ratio ${\cal B}(X\to DD\pi)$, which  is measured experimentally to be larger than $32\%$~\cite{pdg}.  However the total width is poorly known, as $\Gamma_X\lesssim 1.2$~MeV. Using these two  extreme values and the $X\to DD\pi$ decay rate  
\bea
\Gamma(X\to DD\pi)&=&\frac{1}{3}\frac{1}{8\pi m_X^2} 3(g\sqrt{2})^2 p^*(m_X^2,m_D^2,s)\times\notag\\
&\times& \frac{1}{\pi} \frac{s/m_{D^*}\, \Gamma_{D^*}\,{\cal B}(D^*\to D\pi) }{(s-m_{D^{*}}^2)^2+(s/m_{D^*}\, \Gamma_{D^*})^2} \frac{m_{D^*}}{\sqrt{s}} \frac{p^*(s,m_D^2,m_\pi^2)}{p^*(m_{D^*}^2,m_D^2,m_\pi^2)}
\eea
where the decay momentum is $p^*(x,y,z)=\sqrt{\lambda(x,y,z)}/2\sqrt{x}$, $\lambda$ being the K\"all\'en triangular function, it is found that $g\approx 4$~GeV~\cite{brazzi}.

\begin{figure}[ht!]
 \begin{center}
   \includegraphics[width=7truecm]{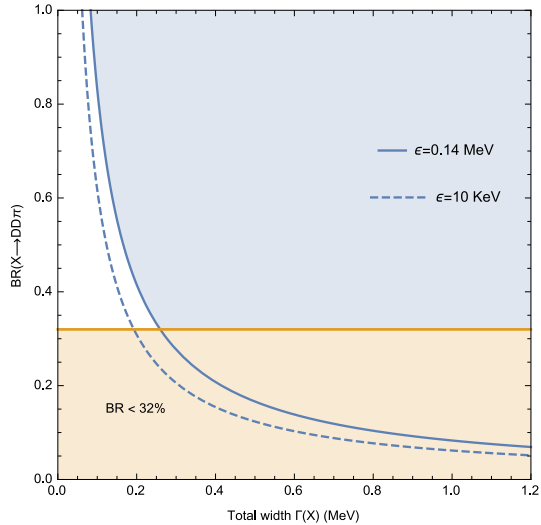}
 \end{center}
\caption{\small Given the experimentally  excluded region (shaded), a loosely bound molecule compatible with~(\ref{lform}), has total width $\Gamma_X$, branching ratio  ${\cal B}(X\to DD\pi)$ and binding energy ${\mathcal E}=m_D+m_{D^*}-m_X$ as in the plot -- here $T\approx -{\mathcal E}\approx 0$. This plot is obtained using the approximate values $m_X=3871.66$~MeV, $m_D=1864.84$~MeV and $m_{D^*}=2006.96$~MeV for ${\mathcal E}=0.14$~MeV and a slightly modified value for $m_X=3871.79$~MeV in order to get ${\mathcal E}=10$~KeV.\label{excluded}}
\end{figure}

Considering for example a branching fraction of ${\cal B}(X\to DD\pi)\simeq 0.32$, we obtain ${\mathcal E}={\mathcal E}_{\rm exp}$  on  assuming a total width of the $X$ as large as $\approx 300$~KeV: lower values of $\Gamma_X$ would also be possible for higher branching ratios ${\cal B}(X\to DD\pi)$, whereas higher  $\Gamma_X$ values are excluded by~(\ref{lform}): see shaded areas in  Fig.~\ref{excluded}.

In all cases we see that only very small kinetic energies are allowed in the $D\bar D^*$ system, definitely smaller than the conservative upper limit on the relative momentum in the center of mass of the pair ($k_{\rm rel}\lesssim 50$~MeV) which was used in~\cite{bigna}. This further suggests that hadronization of prompt $X$ at LHC cannot proceed through a $D\bar D^*$ coalescing phenomenon, the Monte Carlo estimate of the slowly recoiling $D\bar D^*$ being very adverse to the measured high cross section.   

In a deuteron-like description of  $X(3872)$ based on a (three-dimensional) square well potential of depth $U_0\simeq 9$~MeV and range $R\simeq 3$~fm, a bound state energy $B=|{\mathcal E}_{\rm exp}|=0.1$~MeV is obtained. 
The expectation value of the kinetic energy in the bound state $\psi$ is  found to be
\be
\langle T\rangle_\psi\simeq 4~{\rm MeV} \,\;\;\; k_{\rm rel}\simeq 86~{\rm MeV}
\ee
a rather higher value of $T$ with respect to those discussed before. However we have to observe that, in this model, the $D$ and $\bar D^*$ mesons have indeed {\it finite} negative  total energy.
To make this happen, the $D\bar D^*$ pair produced in $pp$ hadronization must interact with at least a third hadron to change its relative kinetic energy and fall  in the discrete (even though superficial) level of the attractive potential. The expected $X$ width would therefore be $\Gamma_X\approx\Gamma_{D^*}\approx 100$~KeV, even at a binding energy as large as ${\mathcal E}=0.14$~MeV (compare to Fig~\ref{excluded}), for it would be  a stable bound state whose lifetime coincides with the lifetime of the shortest lived between its components. Monte Carlo studies on $\pi D\bar D^*$ final state rescatterings do not encourage this picture either~\cite{rescatt}, suggesting overall that hadronization should most likely  produce compact tetraquarks,  which might otherwise be understood in the picture described by~\cite{noi}.

The formation of  $X$ as a loosely $D\bar D^*$ bound state might occur either via a low energy ($T\approx 0$) resonant scattering mechanism (see Fig.~\ref{excluded}) or via multi-body final state interactions producing a deuteron-like state. The latter case, although not supported by Monte Carlo simulations~\cite{rescatt}, might be more realistic when considering $X$ prompt production in high energy $pp$  collisions at the LHC,  with high transverse momentum cuts on hadrons.    

A number of other states are described in the literature as loosely bound hadron molecules: notably the charged $Z_b(10610)$ and $Z_b^\prime(10650)$~\cite{zbs} happen to be very close to $B\bar B^*$ and  $B^*\bar B^*$ thresholds, whereas their analogs in the charm sector, $Z_{c}(3900)$ and $Z_c^\prime(4025)$~\cite{zcs}, are less compelling molecules for their binding energy turns out to be $-{\mathcal E}>0$ by about 20~MeV. Accessing  precise measurements of their properties will enable to further test the relation between binding energy and partial widths as done in Fig.~\ref{excluded} for the $X$. In consideration of the multitude of thresholds that can be formed combining all known open charm and beauty mesons, there are also a number of $Y$ resonances which are eligible molecular candidates.

Although the interpretation of the $X$ in terms of a loosely bound molecule is just given for granted by many, we have to remark that it is challenged by diverse constraints as those discussed in the literature on high energy production~\cite{bigna, rescatt, artobrat} or those implicit in Eq.~(\ref{lform}), and sketched in Fig.~\ref{excluded}. There will hopefully be a number of more precise experimental measurements on the properties of $X,Y,Z$ resonances which will help in disentangling this intricate matter. 

{\bf \emph{Note Added.}} Soon after the preparation of this draft, I noticed the paper by Tomaradze {\it et al.}~\cite{toma} claiming that a precision measurement of the mass difference between $D^0$ and $D^{*0}$ mesons leads to a binding energy of ${\mathcal E}\sim 3\pm192$~KeV  in the $D\bar D^*$ molecule interpretation of $X(3872)$.  

{\bf \emph{Acknowledgements.}}
I wish to thank A. Esposito, J. Ferretti, A. Guerrieri, F. Piccinini and A. Pilloni for fruitful discussions.

\bibliographystyle{unsrt}

\end{document}